# Molecular gas in nearby elliptical radio galaxies

Ocaña Flaquer, B.[*], Leon, S[*], Lim, J.[†], Combes, F.[**] and Dinh-V-Trung[†]

[*]*Instituto de Radio Astronomía milmétrica (IRAM), Avda. Divina PAstora 7, local 20, 18012 Granada, Spain.*
[†]*Academia Sinica, Institute of Astronomy and Astrophysics, P.O. Box 1-87, Nankang, Taipei, 11529, Taiwan*
[**]*Observatoire de Paris, LERMA, 61 Av. de l'Observatoire, F-75014. Paris, France*

**Abstract.** Powerful radio-AGN are hosted by massive elliptical galaxies which are usually very poor in molecular gas. Nevertheless the central Black Hole (BH) needs molecular gas for the nuclear activity. Thus it is important to study the origin, the distribution and the kinematics of the molecular gas in such objects. We have performed at the IRAM-30m telescope a survey of the CO(1-0) and CO(2-1) emission in the most powerful radio galaxies of the Local Universe, selected only on the basis of their radio continuum fluxes. The main result of that survey is the low content in molecular gas of such galaxies compared to Seyfert galaxies. The median value of the molecular gas mass is $4 \times 10^8 M_\odot$. Moreover, the CO spectra indicate the presence of a central molecular gas disk in some of these radio galaxies. We complemented this survey with photometric data of SPITZER and IRAS fluxes with the purpose to study the dust and its relation with the molecular gas and AGN.

**Keywords:** Molecular Hydrogen, molecular spectra, Sky survey, Radio microwave (>1mm), Elliptical galaxies, Active galaxies, ISM, Galaxies.

## SAMPLE AND OBSERVATIONS

The sample contains a total of 52 elliptical radio galaxies selected on the basis of their radio continuum and not on their IR fluxes (eg. Evans et al. 2005 [2]). The sample mainly contains galaxies from the Third Cambridge Catalog (3CR), New General Catalog (NGC) and the Second Bologna Catalog of Radio Sources (B2). There is as well a galaxy (OQ208) from the Ohio State University Radio Survey Catalog which is the most massive galaxy in our sample.

All these galaxies have been observed with the IRAM-30m telescope at the frequencies of the CO(1-0) and CO(2-1) transistors, redshifted to the galaxy velocity. 28 galaxies were detected: out of those galaxies 9 were detected only in CO(1-0), 8 detected in CO(2-1) and 5 detected in both. One of the detected galaxies was detected not only in emission but also in absorption (B2 0116+31). The galaxies 3CR31 and the galaxy 3C264 have been studied in detail previous to our study by Lim et al. 2000 [4] both galaxy show clearly the double horn profile typical of a molecular gas disk, as can be seen in figures 1 and 2. 3C31 was also observed with the IRAM interferometer at the Plateau de Bure in France.

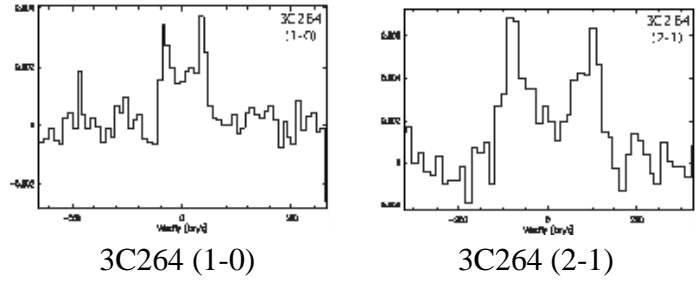

3C264 (1-0)          3C264 (2-1)

**FIGURE 1.** 3C 264, a galaxy clearly detected where the double horn is visible

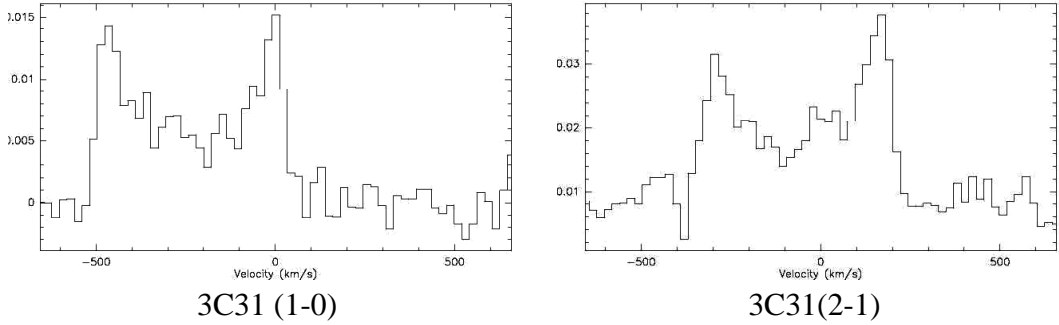

3C31 (1-0)          3C31(2-1)

**FIGURE 2.** CO(1-0) & CO(2-1) Spectra for the galaxy 3C31

## MOLECULAR GAS

We calculated the molecular gas mass in the galaxies using a standard value for the CO-to-$H_2$ conversion factor [3]. The median value is about $4 \times 10^8 M_\odot$ The distribution can be seen in figure 3 where we present a histogram of the mass. Comparing with the survey of Evans et al. 2005 [2] (median value of $M_{H_2}$ is about $7 \times 10^9 M_\odot$) we find a much lower value than for the FIR-selected radio galaxies.

The dust temperature was calculated using the 60 and 100 microns from IRAS data. The dust was modeled by a modified black body where the emissivity is proportional to the frequency. The median value of 35.44 K indicates that the temperature is overall warmer than in normal spiral galaxies, possibly due to the heating of the radio-AGN. The median value of the ratio $H_2$/dust mass is about 250, much lower than the average spiral ( 700).

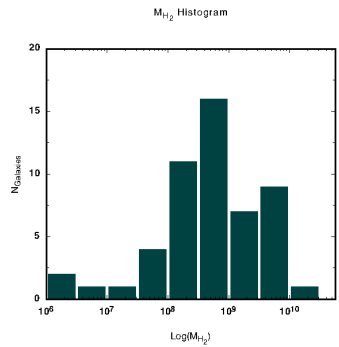

**FIGURE 3.** Mass distribution

There is a clear correlation between the FIR and CO(1-0) emission. In spiral galaxies this correlation arises form the star formation (SF) which is correlated to the moelcular gas content and which heats up the dust. In the elliptical radio galaxies, the star formation is normally at a much lower level. This correlation is linked to the dust heating mechanism (AGN+SF).

The line ratio between the CO(2-1) and the CO(1-0) transitions is computed by comparing the integrated intensity ratio $I_{CO}$(2-1)/$I_{CO}$(1-0). The maximum line ratio was found to be 2.38 and the average value was 1.22. Although the beam size is different at the two frequencies, it is much lower than the line ratio of 0.8 found in the spiral galaxies, indicating a difference in the physical conditions of the molecular gas (T,n) [1]. As an example, look at the spectra of the galaxy 3CR264 (Fig. 1).

## CONCLUSIONS AND FUTURE WORK

Conclusions:

- We have detected molecular gas in 28 of 52 galaxies observed in CO(1-0) and CO(2-1) at the IRAM-30m. Seven of them show a double horn profile spectra, indicating a molecular gas disk.
- The molecular gas mass is lower in our sample ( $10^8 M_\odot$) than in the FIR-selected sample ( $10^9 M_\odot$).
- The CO(2-1)/CO(1-0) line ratio is on average larger than in spiral galaxies with a value of  1.2.
- The dust temperature in the sample is higher than expected, possibly because of the radio-AGN.

Future work:

- Create a dust model of the disk with different temperatures in order to analyze the ISM of the Galaxies.
- Analyze the dynamics of he gas in the very center with the PdBI data.
- Study of the star formation efficiency and the interplay with AGNs
- To analyze the HI content of these radio galaxies to detect possible interactions and to compare with the molecular gas content.